%% file: main.tex
\documentclass[journal]{IEEEtran}
\usepackage{amsmath,amssymb}
\usepackage{graphicx,graphics,color,psfrag}
\usepackage{cite,balance}
\usepackage[labelsep=period]{caption}
\usepackage{framed}
\allowdisplaybreaks
\usepackage{accents}
\usepackage{amsthm}
\usepackage{bm}
\usepackage{url}
\usepackage[english]{babel}
\usepackage{multirow}
\usepackage{enumerate}
\usepackage{cases}
\usepackage{stfloats}
\usepackage{dsfont}
\usepackage{color,soul}
\usepackage{amsfonts}
\usepackage{cite,graphicx,amsmath,amssymb}
\usepackage{fancyhdr}
\usepackage{hhline}
\usepackage{graphicx,graphics}
\usepackage{array,color}
\usepackage{mathtools}
\usepackage{amsmath}
\usepackage[T1]{fontenc}
\usepackage{float}  
\usepackage{subcaption}
\usepackage{stfloats}
\usepackage{siunitx}
\usepackage{algorithm}
\usepackage{algpseudocode}

\include{header}

\begin{document}
\captionsetup[figure]{name={Fig.}}
\title{\huge Flexible Beamforming Design with Two-layer Rotatable Antenna: Synergizing Array and Antenna Rotations}

\author{Weijia Wang, \IEEEmembership{Student Member, IEEE}, Changsheng You, \IEEEmembership{Member, IEEE}, Xiaodan Shao, \IEEEmembership{Member, IEEE}, \\and Rui Zhang, \IEEEmembership{Fellow, IEEE}
\vspace{-15pt}

\thanks{
\vspace{-6pt}

Weijia Wang is with School of Science and Engineering, The Chinese University of Hong Kong, Shenzhen, Guangdong 518172, China (e-mail: weijiawang@link.cuhk.edu.cn).

Changsheng You is with the Department of Electronic and Electrical Engineering, Southern University of Science and Technology (SUSTech), Shenzhen 518055, China (e-mail: youcs@sustech.edu.cn).

 Xiaodan Shao is with the Department of Electrical and Computer Engi
neering, University of Waterloo, Waterloo, ON N2L 3G1, Canada (email:
 x6shao@uwaterloo.ca).

Rui Zhang is with the Department of Electrical and Computer Engineering, National University of Singapore, Singapore 117583 (e-mail: elezhang@nus.edu.sg).

\textit{Corresponding author: Changsheng You and Xiaodan Shao.}
}}

\setlength{\abovedisplayskip}{4pt} 
\setlength{\belowdisplayskip}{4pt} 

\everymath{%
  \thinmuskip=0.5mu   
  \medmuskip=0.5mu    
  \thickmuskip=0.5mu  
}

\maketitle
\begin{abstract}
Reconfigurable antenna technology, such as movable antennas (MAs) and rotatable antennas (RAs), has emerged as a promising solution to enhance wireless communication performance by exploiting new degrees of freedom (DoFs) in antenna reconfiguration. However, existing RA designs mostly consider array-wise or antenna-wise rotation only, limiting their potential in wide-range radiation pattern control. To overcome this, we propose a new two-layer RA architecture for downlink coverage improvement, which combines array-wise rotation for global orientation adjustment with per-antenna rotation for fine-grained radiation refinement. We then formulate an optimization problem to maximize the minimum beamforming gain over a target region by jointly optimizing the two-layer rotations and transmit beamforming. To solve this non-convex problem, an efficient block coordinate descent (BCD) algorithm is proposed, which alternately optimizes one of the three variables in an iterative manner, with the other two being fixed. Numerical results demonstrate that the proposed two-layer RA significantly improves the minimum beamforming gain over fixed antenna arrays and single-layer RAs.
\end{abstract}


\vspace{-10pt}
\section{Introduction}
\vspace{-6pt}
Next-generation wireless networks are expected to support substantially higher capacity and data rates with stringent latency requirements\cite{11329408}. Meeting these demands calls for highly efficient spatial resource utilization, which is commonly realized via directional beamforming. However, conventional fixed-position antenna arrays (FPAs) employ fixed array geometry and radiation characteristics, which fundamentally constrain the communication performance in dynamic wireless environments.

Movable antennas (MAs) and fluid antenna systems (FASs) have recently attracted growing attention as they provide new spatial degrees of freedom (DoFs) for wireless system designs \cite{Millimeter-wave, 9264694}. By flexibly adjusting antenna positions within a prescribed region, these architectures can dynamically reshape wireless channels to improve capacity without increasing the number of antennas. However, the lack of rotation-related DoFs limits their efficacy in tailoring array radiation patterns to complex spatial environments. To fully exploit all six-dimensional (6D) spatial DoFs, 6D movable antenna (6DMA) has been proposed to jointly adjust the three-dimensional (3D) position and 3D rotation of distributed antennas/arrays \cite{10752873}. As a cost-effective implementation of 6DMA, rotatable antennas (RAs) have been recently proposed \cite{11222668}, which enhance wireless system performance with less hardware/implementation cost than 6DMA. Nevertheless, existing RA architectures typically adopt either antenna-wise or array-wise orientation control. The former offers flexibility but is subject to physical rotation constraints \cite{10906511}, while the latter controls overall array orientation but lacks fine-grained per-antenna radiation configuration \cite{11039664}.

To overcome these limitations, we propose in this letter a new two-layer RA architecture that synergizes the array rotation for global orientation adjustment and per-antenna rotation for fine-grained radiation refinement, hence offering great flexibility in beam pattern shaping. Specifically, we formulate an optimization problem to maximize the minimum beamforming gain by jointly optimizing the transmit beamforming and two-layer rotation angles. To solve this non-convex problem, we propose an efficient block coordinate descent (BCD) algorithm, which alternately optimizes one of the three variables in an iterative manner, with the other two being fixed. Numerical results demonstrate notable worst-case beamforming gain improvements of the proposed architecture over conventional fixed and single-layer RAs.

  \begin{figure}[!t]
\centering
\includegraphics[width=0.32\textwidth]{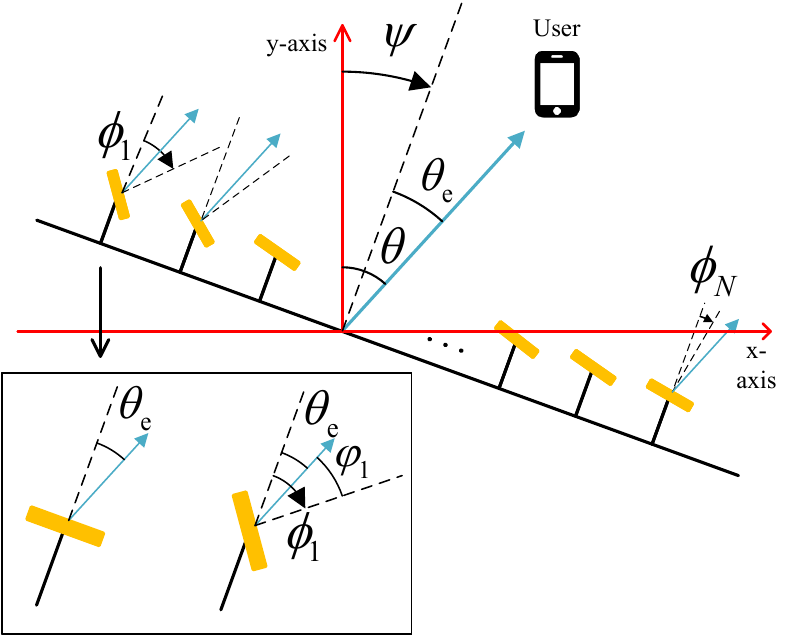}
\caption{Proposed two-layer RA for downlink communication.    } 
\label{system:model} 
\vspace{-15pt}
\end{figure}

\vspace{-10pt}
\section{System Model and Problem Formulation} 
\label{Sec:System model}
\vspace{-6pt}
\subsection{System Model}
\label{sec:system_model}
We consider a narrowband downlink communication system as shown in Fig.~1, where a base station (BS) serves a target coverage region via a reconfigurable uniform linear array (ULA) \footnote{The results of this letter apply naturally to uplink communications with similar user coverage requirement.}. The ULA comprises $N$ directional antennas, denoted by $\mathcal{N}\triangleq\{1,2,\ldots,N\}$, with an inter-antenna spacing of $d_0 = \lambda/2$, where $\lambda$ denotes the carrier wavelength. The aperture size of the antenna is $\sqrt{A}\times\sqrt{A}$ with $\sqrt{A}\leq d_0$. 

\underline{\bf Two-layer rotation model:} For the proposed two-layer RA architecture, the array-wise rotation mechanically changes the orientation of the entire ULA, while the antenna-wise rotation adjusts individual antenna boresights, thereby effectively adjusting its radiation pattern \footnote{In practice, two-layer RA can be realized via mechanical actuators (e.g., servo motors or micro-electromechanical systems) with a wide angular range and millisecond-scale response time~\cite{11222668}. The non-negligible response time makes two-layer RA best suited to quasi-stationary scenarios where the beam coverage region changes much more slower than the mechanical response~\cite{10906511}.}.

Specifically, we consider a two-dimensional global Cartesian coordinate system (GCS) with the origin located at the center of ULA. As shown in Fig.~1, the ULA is rotated by an angle $\psi$, with clockwise rotation defined as the positive direction. The array rotation is constrained by $|\psi| \le \psi_{\max}$, where $\psi_{\max}$ is the maximum allowable array rotation angle. Consequently, the coordinates of the $n$-th antenna are given by
\begin{equation}
    \mathbf{p}_n(\psi) = \frac{2n-N-1}{2} d_0 \left[ \cos\psi, \, \sin\psi \right]^T, ~ \forall n\in \mathcal{N}.
\end{equation}
Next, to enable fine-grained beam control, each antenna controls the boresight of its radiation pattern. Let $|\phi_n|$ denote the rotation angle of the $n$-th antenna, which is defined as the angle with respect to (w.r.t.) the array normal in the same positive direction, and $\boldsymbol{\phi} \triangleq [\phi_{1},\phi_{2},\ldots,\phi_{N}]^{T}\in \mathbb{R}^{N\times 1}$ as the set of rotation angles of all antennas. In practice, large antenna rotation angles may incur mutual coupling and distort radiation patterns \cite{11222668}. To mitigate mutual coupling effects, the per-antenna rotation is constrained by $|\phi_n| \le \phi_{\max},~ \forall n\in \mathcal{N}$.

Let $G(\varphi)$ denote the directional gain of each rotatable antenna, which is a function of the angle between the incident direction and the antenna boresight, denoted as $\varphi$ (see Fig.~1). Similar to \cite{Zheng2026RA}, we adopt a unified cosine-power gain model, where the antenna directional gain is given by
\begin{equation}
    G(\varphi) \triangleq
    \begin{cases}
        G_{\max} \cos^{2p}(\varphi), & |\varphi| \le \frac{\pi}{2}, \\
        0, & \text{otherwise}.
    \end{cases}
    \label{eq:unified_gain}
\end{equation}
Herein, $G_{\max}=2(2p+1)$ is the maximum directional gain achieved at the boresight angle (i.e., $\varphi = 0$) for satisfying power conservation, and $p \ge 1/2$ is the directivity factor that determines the antenna beamwidth.

\underline{\bf Channel model:} For beam coverage design, we consider a typical single-antenna user in the target region, whose angle of departure (AoD) w.r.t. GCS is denoted by $\theta$. Let $\mathbf{h}^{H} \in \mathbb{C}^{1 \times N}$ denote the downlink channel from the BS to the typical user. For the considered scenarios of quasi-stationary line-of-sight (LoS) channels, $\mathbf{h}$ can be modeled as
\begin{equation}
    \mathbf{h}^{H} = h \mathbf{a}^{H}(\theta;~ \psi, \boldsymbol{\phi}),
\end{equation}
where $h=\rho (r_{0}/r)^{\gamma}e^{-j 2\pi r/\lambda}$ is the complex-valued path gain at distance $r$, with $\rho$ denoting the reference channel gain at a distance $r_{0}=1$ meter (m) and $\gamma$ being the path-loss exponent. In addition, the array response vector is given by
\begin{equation}
    \mathbf{a}^{H}(\theta;~ \psi, \boldsymbol{\phi}) = \mathbf{v}^{H}(\theta; \psi)\mathbf{D}(\theta;~ \psi, \boldsymbol{\phi}),
\end{equation}
where $\mathbf{v}^{H}(\theta;~ \psi) = [ 1, e^{-j k d_0\sin\theta_{\rm e}}, \ldots, e^{-j k d_0 (N-1)\sin\theta_{\rm e}} ]$ represents the channel steering vector with $\theta_{\rm e} \triangleq \theta - \psi$ being the effective AoD of the user. $\mathbf{D}(\theta;\psi,\boldsymbol{\phi}) = \text{diag}(\sqrt{G(\varphi_{1})}, \ldots, \sqrt{G(\varphi_{N})})$ denotes the antenna directional gains of all antennas, with $G(\varphi_{n})$ being that of the $n$-th antenna, and $\varphi_{n}=\theta_{\rm e}-\phi_{n}$ denoting the incident angle from the user to the $n$-th antenna.

To reduce hardware complexity, we consider analog beamforming-based beam coverage, where $\mathbf{w} \in \mathbb{C}^{N \times 1}$ with $|w_n| = 1/\sqrt{N}$ is the analog beamformer of the ULA \footnote{Practical phase shifters with finite phase resolution introduce quantization errors that cause the realized beamforming gain to deviate from the continuous-phase solution. Our continuous-phase formulation thus provides an upper bound on achievable performance.}. The received signal power at the typical user is given by
\begin{equation}
    P_{\rm r} = P_{\rm t} |h|^2 \left| \mathbf{a}^{H}(\theta;~ \psi, \boldsymbol{\phi}) \mathbf{w} \right|^2,
    \label{eq:received_power}
\end{equation}
where $P_{\rm t}$ is the BS transmit power. We define $G^{\rm b}(\theta;~\psi,\boldsymbol{\phi},\mathbf w) \triangleq \bigl|\mathbf a^H(\theta;~\psi,\boldsymbol{\phi})\,\mathbf w\bigr|^2$ as the (normalized) beamforming gain at the user along direction $\theta$.

\begin{remark} [Rician Fading Channel Model]
    \emph{This work can be extended to Rician fading scenarios by optimizing the expected beamforming gain \cite{multipath}. The Rician fading channel can be expressed as
    $\mathbf{h}^{H} = \sqrt{\frac{K}{K+1}} h_{\rm LoS} \mathbf{a}^{H}(\theta; \psi, \boldsymbol{\phi}) + \sqrt{\frac{1}{K+1}} h_{\rm NLoS} \mathcal{CN}(\mathbf{0}, \mathbf{I}_N)$,
    where $K$ is the Rician factor, $h_{\rm LoS} = \rho(r_{0}/r)^{\gamma} e^{-j 2\pi r/\lambda}$ is the complex-valued LoS path gain, and $h_{\rm NLoS} = \rho(r_{0}/r)^{\gamma}$ is the non-LoS (NLoS) path gain. The expected received signal power can then be obtained as
    $\bar{P}_{\rm r} = P_{\rm t}  \rho^{2}(r_{0}/r)^{2\gamma}(\theta; \psi, \boldsymbol{\phi}, \mathbf{w})$,
    where $\bar{G}^{\rm b}(\theta; \psi, \boldsymbol{\phi}, \mathbf{w}) \triangleq \frac{K}{K+1} |\mathbf{a}^{H}(\theta; \psi, \boldsymbol{\phi}) \mathbf{w}|^2 + \frac{1}{K+1}$ is defined as the expected beamforming gain.}
\end{remark}

\vspace{-12pt}
\subsection{Problem Formulation}
\vspace{-4pt}
Let $\Theta$ denote the target coverage region, which generally consists of $M$ disjoint angular intervals, modeled as
\begin{align}
    \Theta \triangleq \bigcup_{m=1}^{M}[\alpha_m,~\beta_m],
\end{align}
where $-\tfrac{\pi}{2}\le \alpha_1 < \beta_1 < \cdots < \alpha_M < \beta_M \le \tfrac{\pi}{2}$. Our goal is to maximize the minimum (normalized) beamforming gain over the entire region, denoted as $G^{\rm b}_{\rm worst}(\psi,\boldsymbol{\phi},\mathbf w) \triangleq \min_{\theta\in\Theta} G^{\rm b}(\theta;~\psi,\boldsymbol{\phi},\mathbf w)$, via joint optimization of array rotation $\psi$, per-antenna rotation $\boldsymbol{\phi}$, and the analog beamformer $\mathbf{w}$. Based on the above, this optimization problem can be mathematically formulated as
\begin{subequations} \label{prob:P1}
    \begin{align}
    \text{\textbf{(P1})}: \quad
    \max_{\psi,~\boldsymbol{\phi},~\mathbf w} \quad
    &G^{\rm b}_{\rm worst}(\psi,\boldsymbol{\phi},\mathbf w)=\min_{\theta\in\Theta}\,
    \bigl|\mathbf a^H(\theta;~\psi,\boldsymbol{\phi})\,\mathbf w\bigr|^2\nonumber \\
    \text{s.t.} \quad
    & |w_n|=\tfrac1{\sqrt N}, \quad \forall n\in\mathcal{N}, \label{const:modulus}\\
    & |\phi_n|\le\phi_{\max}, \quad \forall n\in\mathcal{N}, \label{const:ant_rot}\\
    & |\psi| \le \psi_{\max}, \label{const:array_rot}
    \end{align}
\end{subequations}
where \eqref{const:modulus} imposes the unit modulus constraint for analog beamforming, \eqref{const:ant_rot} limits per-antenna rotation in a finite regime to mitigate mutual coupling, and \eqref{const:array_rot} limits array rotation in a finite regime.

\vspace{-10pt}
\section{Proposed Algorithm for Solving (\textbf{P1})}
\vspace{-4pt}
In this section, we develop an efficient algorithm to obtain a high-quality suboptimal solution to problem~(\textbf{P1}). Problem (\textbf{P1}) is non-convex and generally difficult to solve optimally, due to the intricately coupled variables and non-convex constraint \eqref{const:modulus}. To tackle this challenge, we adopt an BCD framework, which alternately optimizes one of
the three variables in an iterative manner, with the other two being fixed.

\vspace{-10pt}
\subsection{Problem Reformulation}
\vspace{-4pt}
Problem (\textbf{P1}) is a joint optimization over the array rotation $\psi$, the per-antenna rotation $\boldsymbol{\phi}$, and the beamforming vector $\mathbf{w}$. To solve it, we first discretize the target angular region $\Theta$. Since $\Theta$ consists of $M$ disjoint angular intervals, each sub-interval $\Theta_m = [\alpha_m, \beta_m]$ is quantized into $Q_m$ angular samples, denoted by $\theta_{m,q} = \alpha_m + \frac{q-1}{Q_m-1}(\beta_m - \alpha_m),~\forall q \in \{1, \dots, Q_m\}$. Let $\{\theta_q\}_{q \in \mathcal{Q}}$ denote the set of all $Q = \sum_{m=1}^{M} Q_m$ sampling points, where $\mathcal{Q} \triangleq \{1, \dots, Q\}$. Then, by introducing an auxiliary variable $\tau$, problem (\textbf{P1}) is reformulated as
\begin{subequations}\label{eq:P2-reformulated}
\begin{align}
\mathrm{(\textbf{P2}):}\quad 
&\max_{\psi,\,\boldsymbol{\phi},\,\mathbf{w},\,\tau}\quad \tau\\
\text{s.t.}\quad 
&\left|\mathbf{a}^H(\theta_q;~\psi,\boldsymbol{\phi})\,\mathbf{w}\right|^2 ~\ge~ \tau,
\quad \forall q\in \mathcal{Q},\\
&|w_n|=\frac{1}{\sqrt{N}},\quad \forall n\in \mathcal{N},\\
&|\phi_n|\le\phi_{\max},\quad \forall n\in \mathcal{N},\\
&|\psi|\le\psi_{\max}.
\end{align}
\end{subequations}
Since $\psi$, $\boldsymbol{\phi}$, and $\mathbf{w}$ are mutually coupled in the steering vector $\mathbf{a}(\theta_q;~\psi,\boldsymbol{\phi})$ and the constraint is non-convex with respect to each variable, a direct joint optimization is intractable. We therefore adopt an BCD framework, where two of the three variables are fixed while the remaining one is optimized, and this process is iterated until convergence.

\vspace{-8pt}
\subsection{Per-antenna Rotation Optimization}
Given any feasible $\mathbf{w}$ and $\psi$, problem (\textbf{P2}) reduces to
\begin{align}
(\textbf{P2.1}):\quad
\max_{\boldsymbol{\phi},\,\tau}\quad & \tau \nonumber\\
\mathrm{s.t.}\quad 
& \left|\mathbf{a}^{H}(\theta_q;~\psi,\boldsymbol{\phi})\mathbf{w}\right|^2 \ge \tau,\quad \forall q\in\mathcal{Q},\nonumber\\
& |\phi_n|\le \phi_{\max},\quad \forall n \in \mathcal{N}.\nonumber
\end{align}
Let $g_q(\boldsymbol{\phi}) \triangleq \left|\mathbf{a}^{H}(\theta_q;~\psi,\boldsymbol{\phi})\mathbf{w}\right|^2$ denote the beamforming gain at the AoD of $\theta_{q}$. Since problem $(\textbf{P2.1})$ is a non-convex problem due to the non-concave function $g_q(\boldsymbol{\phi})$ w.r.t. $\boldsymbol{\phi}$, the optimal solution of problem $(\textbf{P2.1})$ is difficult to obtain. To address this issue, we apply the successive convex approximation (SCA) technique to obtain a suboptimal solution to problem (\textbf{P2.1}) in an iterative manner. Without loss of generality, we present the procedure of the ($i+1$)-th iteration and denote the solutions of $\boldsymbol{\phi}$ and $\tau$ obtained in the $i$-th iteration by $\boldsymbol{\phi}^{(i)}$ and $\tau^{(i)}$, respectively. Specifically, for the non-concave function $g_q(\boldsymbol{\phi})$, we construct the following linear surrogate function $\tilde{g}_q(\boldsymbol{\phi}|\boldsymbol{\phi}^{(i)})$ to approximate $g_q(\boldsymbol{\phi})$ by applying the first-order Taylor expansion at $\boldsymbol{\phi}^{(i)}$,
\begin{align}
g_q(\boldsymbol{\phi})\geq\tilde{g}_q(\boldsymbol{\phi}|\boldsymbol{\phi}^{(i)})
\triangleq&
g_q(\boldsymbol{\phi}^{(i)})
+ \nabla_{\boldsymbol{\phi}} g_q(\boldsymbol{\phi}^{(i)})^{T}
(\boldsymbol{\phi}-\boldsymbol{\phi}^{(i)}).
\end{align}
Herein, $\nabla_{\boldsymbol{\phi}} g_q(\boldsymbol{\phi}^{(i)})$ represents the gradient vector of $g_q(\boldsymbol{\phi})$ evaluated at $\boldsymbol{\phi}^{(i)}$ with its elements given  by
\begin{align}
&\left[\nabla_{\boldsymbol{\phi}} g_q(\boldsymbol{\phi}^{(i)})\right]_n
=\nonumber
\\& 2\,\mathrm{Re}\left\{ 
\mathbf{w}^{H} \mathbf{a}(\theta_q;~\psi,\boldsymbol{\phi}^{(i)})
\frac{\partial \mathbf{a}^{H}(\theta_q;~\psi,\boldsymbol{\phi}^{(i)})}{\partial \phi_n} \mathbf{w} 
\right\}, ~\forall n\in\mathcal{N}.\nonumber
\end{align}

By replacing $g_q(\boldsymbol{\phi})$ in $g_q(\boldsymbol{\phi})\geq \tau$ with its lower bound $\tilde{g}_q(\boldsymbol{\phi}|\boldsymbol{\phi}^{(i)})$ in (10), problem $(\textbf{P2.1})$ can be approximated as
\begin{align}
(\textbf{P2.2}):\quad
\max_{\boldsymbol{\phi},\,\tau}\quad & \tau \nonumber\\
\mathrm{s.t.}\quad 
& \tilde{g}_q(\boldsymbol{\phi}|\boldsymbol{\phi}^{(i)}) \ge \tau,\quad \forall q\in \mathcal{Q},\nonumber \\
& |\phi_n|\le \phi_{\max},\quad \forall n \in \mathcal{N}.\nonumber
\end{align}
Problem (\textbf{P2.2}) is a convex optimization problem with affine constraints, which can be efficiently solved using standard solvers, e.g., CVX solver.

\vspace{-8pt}
\subsection{Array Rotation Optimization}
Given any feasible per-antenna rotation $\boldsymbol{\phi}$ and $\mathbf{w}$, problem (\textbf{P2}) reduces to
        \begin{align}
        (\textbf{P2.3}):\quad
        \max_{\psi,\,\tau}\quad & \tau \nonumber\\
        \mathrm{s.t.}\quad
        & \left|\mathbf{a}^{H}(\theta_q;~\psi,\boldsymbol{\phi})\mathbf{w}\right|^2 \ge \tau,\quad \forall q\in\mathcal{Q},\nonumber\\
        & |\psi|\le \psi_{\max}.\nonumber
        \end{align}
        Let $f_q(\psi) \triangleq \left|\mathbf{a}^{H}(\theta_q;~\psi,\boldsymbol{\phi})\mathbf{w}\right|^2$. The epigraph form of (\textbf{P2.3}) is equivalent to maximizing $F(\psi) \triangleq \min_{q\in\mathcal{Q}} f_q(\psi)$, which is non-smooth and thus cannot be directly solved by gradient-based methods. We apply the log-sum-exp smoothing approximation \cite{Palomar2006}
        \begin{equation}
        F(\psi)\approx F_\mu(\psi) \triangleq -\frac{1}{\mu} \ln \sum_{q \in \mathcal{Q}} e^{-\mu f_q(\psi)},
        \end{equation}
        where $\mu > 0$ is a smoothing parameter satisfying $F_\mu(\psi) \le F(\psi) \le F_\mu(\psi) + \frac{\ln Q}{\mu}$. Problem (\textbf{P2.3}) is then approximated as
        \begin{align}
        (\textbf{P2.4}):\quad
        \max_{\psi}\quad & F_\mu(\psi) \nonumber\\
        \mathrm{s.t.}\quad
        & |\psi|\le \psi_{\max},\nonumber
        \end{align}
        which is convex since $F_\mu(\psi)$ is concave. We solve (\textbf{P2.4}) via the projected gradient ascent (PGA). Specifically, the gradient of $F_\mu$ is
        \begin{equation}
        \frac{d F_\mu}{d\psi} = \sum_{q \in \mathcal{Q}} \omega_q(\psi) \cdot \frac{d f_q(\psi)}{d\psi},
        \end{equation}
        where $\omega_q(\psi) = \frac{e^{-\mu f_q(\psi)}}{\sum_{q'} e^{-\mu f_{q'}(\psi)}}$ and
        $\frac{df_q}{d\psi} = 2\operatorname{Re}\!\left[ s_q^H(\psi)\, \frac{d s_q}{d\psi} \right]$
        with $s_q(\psi) \triangleq \mathbf{a}^H(\theta_q;~\psi,\boldsymbol{\phi})\mathbf{w}$.
        Substituting $w_n = \frac{1}{\sqrt{N}}e^{j\arg(w_n)}$
        and $[\mathbf{a}^H]_n = \sqrt{G_{\max}}\cos^p(\theta_q-\psi-\phi_n)\,e^{-jkd_0(n-1)\sin(\theta_q-\psi)}$ yields $s_q(\psi) = \sum_{n=1}^{N}\alpha_n(\psi)\,e^{j\beta_n(\psi)}$, where $\alpha_n(\psi) \triangleq \frac{\sqrt{G_{\max}}}{\sqrt{N}}\cos^p(\theta_q-\psi-\phi_n)$
        and $\beta_n(\psi) \triangleq \arg(w_n)-kd_0(n-1)\sin(\theta_q-\psi)$. The derivative is
        \begin{equation}
        \frac{d s_q}{d\psi} = \sum_{n=1}^{N} \left( \frac{d\alpha_n}{d\psi}
        + j\alpha_n \frac{d\beta_n}{d\psi} \right) e^{j\beta_n(\psi)},
        \end{equation}
        where $\frac{d\alpha_n}{d\psi} = \frac{p\sqrt{G_{\max}}}{\sqrt{N}}\cos^{p-1}(\theta_q-\psi-\phi_n)\sin(\theta_q-\psi-\phi_n)$ and $\frac{d\beta_n}{d\psi} = kd_0(n-1)\cos(\theta_q-\psi)$.
        The PGA update with Armijo line search is
        \begin{equation}
        \psi^{(t+1)} = \mathcal{P}_{[-\psi_{\max},\psi_{\max}]}\!\left(\psi^{(t)} + \epsilon^{(t)} \frac{d F_\mu}{d\psi}\bigg|_{\psi^{(t)}}\right),
        \end{equation}
        where $\mathcal{P}_{[-\psi_{\max},\psi_{\max}]}(x) = \min(\psi_{\max}, \max(-\psi_{\max}, x))$. To reduce the risk of local optimum, we initialize from $L$ uniformly spaced points $\psi_\ell^{(0)} = -\psi_{\max} + \frac{2(L-1)\psi_{\max}}{L-1}$ and retain the solution with the largest $F(\psi)$. The smoothing parameter is updated via $\mu^{(t+1)} = \rho\mu^{(t)}$ with $\rho > 1$ to progressively tighten the approximation.

\vspace{-6pt}
\subsection{Transmit Beamforming Optimization}
Given any feasible per-antenna rotation $\boldsymbol{\phi}$ and array rotation $\psi$, problem (\textbf{P2}) reduces to the following problem for transmit beamforming optimization, i.e.,
\begin{subequations}  
\begin{align}  
(\textbf{P2.5}):\max_{\mathbf{w},\,\tau}\quad &\tau\\  
\mathrm{s.t.}\quad  
&\left|\mathbf{a}^H(\theta_q;~\psi,\boldsymbol{\phi})\,\mathbf{w}\right|^2\ge \tau,~\forall q \in\mathcal{Q},\\  
&|w_n|=\frac{1}{\sqrt{N}},~\forall n\in \mathcal{N}. \label{eq:cm_const_refined}  
\end{align}  
\end{subequations}  
To effectively handle the non-convex constant-modulus constraint $|w_n|=1/\sqrt{N}$, we employ the semidefinite relaxation (SDR) technique by defining $\mathbf{W}=\mathbf{w}\mathbf{w}^H$, where $\mathbf{W} \succeq \mathbf{0}$ and $\text{rank}(\mathbf{W})=1$. Then, we introduce a log-determinant penalty to promote a rank-one solution \cite{penalty}. Let 
$\mathbf{A}_q = \mathbf{a}(\theta_q;~ \psi, \boldsymbol{\phi}) \mathbf{a}^H(\theta_q;~ \psi, \boldsymbol{\phi})$. Problem (\textbf{P2.5}) can be rewritten as follows based on the penalty-based SDR technique,
\begin{subequations}  
\begin{align}  
(\textbf{P2.6}):\max_{\mathbf{W}\succeq \mathbf{0},\,\tau}\quad &\tau - \eta \ln\det(\mathbf{W}+\zeta\mathbf{I})\\  
\mathrm{s.t.}\quad &\mathrm{Tr}(\mathbf{A}_q\,\mathbf{W})\ge \tau,~\forall q \in\mathcal{Q},\\  
&[\mathbf{W}]_{n,n}=\frac{1}{N},~\forall n\in\mathcal{N}, \label{eq:relaxed_cm_k}
\end{align}  
\end{subequations}
where $\zeta>0$ is set as a small positive constant to ensure $\det(\mathbf{W}+\zeta\mathbf{I})>0$.
Starting from a small $\eta^{(0)}$, we solve (\textbf{P2.6}) and update the penalty at the $j$-th iteration as 
$\eta^{(j+1)}=\kappa \eta^{(j)}$ with $\kappa>1$ until 
$\lambda_{\max}(\mathbf{W}^\star)/\mathrm{Tr}(\mathbf{W}^\star)\ge 1-\delta$.
Since (\textbf{P2.6}) is a convex problem, the subproblem in each iteration can be solved optimally via interior-point methods. 
Finally, we recover $\mathbf{w}^\star$ as $\mathbf{w}^\star=\frac{1}{\sqrt{N}}\exp\!\left(j\arg(\mathbf{v}_{\max})\right)$, where $\mathbf{v}_{\max}$ is the principal eigenvector of $\mathbf{W}^\star$. To ensure monotonicity, if the extracted $\mathbf{w}^\star$ decreases the objective, we retain the previous solution.

 \begin{remark}[Initialization Strategy]\label{Initialization}
     \emph{To handle the non-convexity of the BCD algorithm, we initialize per-antenna rotation by steering all antenna boresights toward the center of the target region $\Theta$, and array rotation $\psi$ is set to $0^{\circ}$. Specifically, let $\bar{\theta} \triangleq \frac{1}{M}\sum_{m=1}^{M}\bar{\theta}_m$ denote the center of region $\Theta$, where $\bar{\theta}_m$ is the center of the $m$-th angular interval. Then, we initialize the per-antenna rotation as}
    \begin{equation}\label{eq:varphi_init}
        \phi_n^{(0)} = \bar{\theta} , \quad \forall n\in\mathcal{N},
    \end{equation}
    \emph{which provides a starting point for subsequent optimization.}
 \end{remark}

\begin{remark}[Algorithm Convergence and Computational Complexity]
\emph{The BCD algorithm converges since the objective is upper bounded by $NG_{\max}$ and monotonically non-decreasing in each subproblem iteration. The computational complexities of the SCA-based subproblem for $\boldsymbol{\phi}$, the PGA-based subproblem for $\psi$, and the penalty-based SDP for $\mathbf{w}$ are $\mathcal{O}\!\left(I_{\rm SCA}(N+1)^3\sqrt{Q+2N}\right)$, $\mathcal{O}\!\left(L I_{\rm PGA} QN\right)$, and $\mathcal{O}\!\left(I_{\rm SDR}(N^6+QN^4)\right)$, respectively, yielding an overall complexity of $\mathcal{O}\!(I_{\rm BCD}[I_{\rm SCA}(N+1)^3\sqrt{Q+2N}+L I_{\rm PGA} QN+I_{\rm SDR}(N^6+QN^4)])$, where $I_{\rm BCD}$, $I_{\rm SCA}$, $I_{\rm PGA}$, and $I_{\rm SDR}$ denote the iteration numbers of BCD, SCA, PGA, and penalty-based SDR, respectively.}
\end{remark}

\section{Numerical Results}
In this section, numerical results are presented to validate the performance gain of the proposed two-layer RA architecture. Unless otherwise specified, the system parameters are set in Table I. The convergence thresholds for the BCD algorithm, SCA, PGA, and SDR method are set as $10^{-5}$, $10^{-4}$, $10^{-4}$, and $10^{-4}$, respectively. 

\begin{table}[!t]
\centering
\caption{Simulation Parameters}
\label{tab:system_parameters}
\begin{tabular}{ll|ll}
\hline
\textbf{Parameter} & \textbf{Value} & \textbf{Parameters} & \textbf{Value} \\
\hline
 $N$                   & $10$             & $d_0$                & $\lambda/2$                  \\
  $\gamma$                & $2.5$             & $G_{\max}$           & $4$                 \\
$Q$                   & $1000$          &  $p$                   & $1$   \\
$\psi_{\max}$         & $\pi/3$ rad                 & $\phi_{\max}$     & $\pi/3$ rad              \\
$\Delta \psi$         & $1^{\circ}$            & $L$                  & $10$       \\
$\zeta$            & $10^{-6}$          &$\kappa$             & $1.2$        \\
$\delta$              & $10^{-4}$                   &$\eta^{(0)}$         & $10^{-3}$            \\
\hline
\end{tabular}
\end{table}

\begin{figure}[!t]
 \centering
 \begin{subfigure}[t]{0.24\textwidth}
   \centering
   \includegraphics[width=\textwidth]{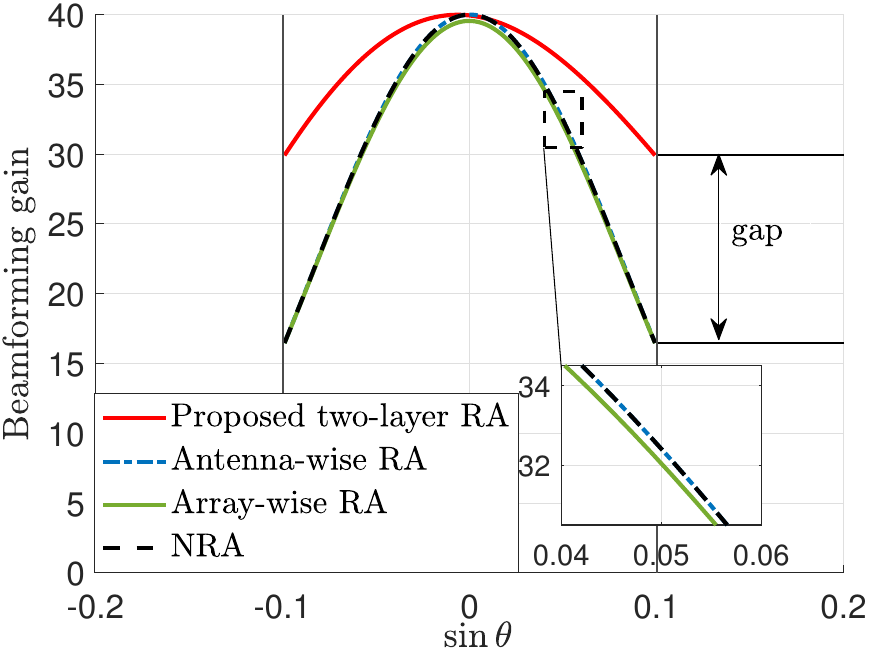}
   \caption{Target region $[-0.1,0.1]$}
   \label{fig:2a}
 \end{subfigure}%
 \hfill
 \begin{subfigure}[t]{0.24\textwidth}
   \centering
   \includegraphics[width=\textwidth]{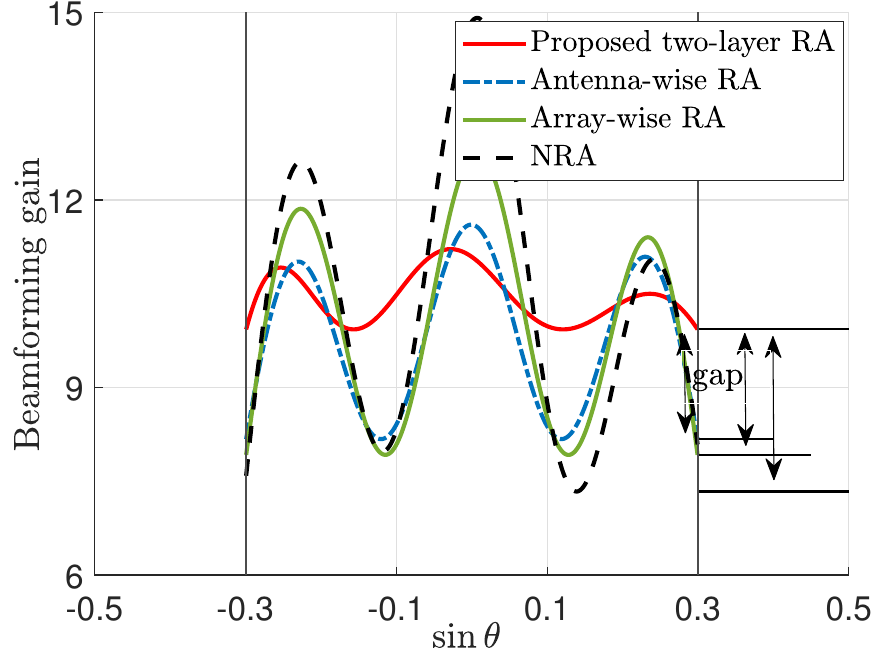}
   \caption{Target region $[-0.3,0.3]$}
   \label{fig:2c}
 \end{subfigure}%
 \hfill
 \vspace{1em} 
 
 \begin{subfigure}[t]{0.24\textwidth}
   \centering
   \includegraphics[width=\textwidth]{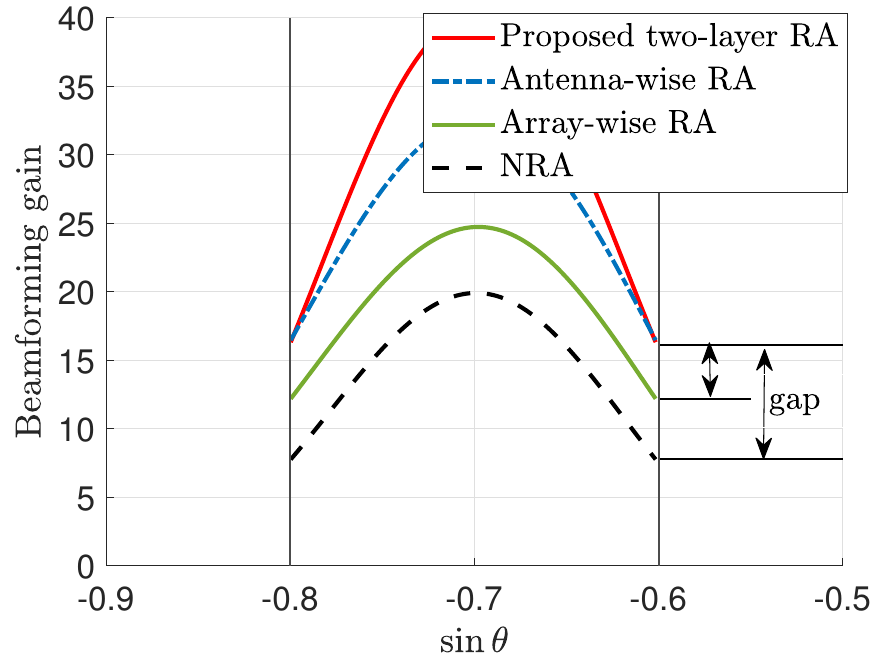}
   \caption{Target region $[-0.8,-0.6]$}
   \label{fig:2b}
 \end{subfigure}%
 \hfill
 \begin{subfigure}[t]{0.24\textwidth}
   \centering
   \includegraphics[width=\textwidth]{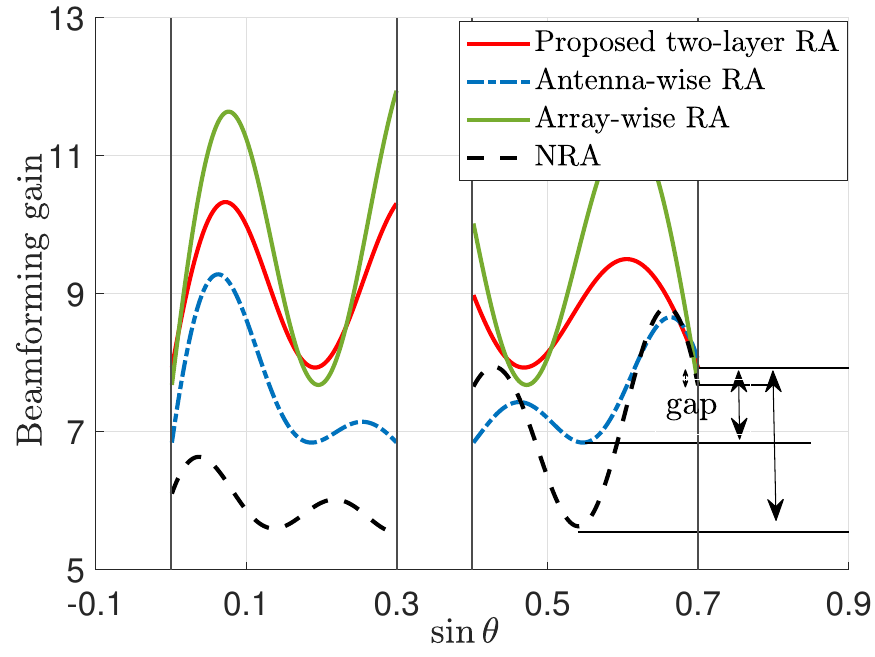}
   \caption{Multi-region coverage}
   \label{fig:2d}
 \end{subfigure}%
 \caption{Beam pattern comparison under different spatial coverage scenarios.}
 \label{fig:beam-patterns}
 \vspace{-6pt}
\end{figure}

For performance comparison, we consider the following benchmark array architectures with a 10-antenna ULA: 1) \textbf{Antenna-wise RA}, where only the antenna-wise rotation is considered; 2) \textbf{Array-wise RA}, where only the array-wise rotation is considered; and 3) \textbf{Non-RA (NRA)}, which represents a fixed-rotation ULA without any rotation control. Moreover, to evaluate the performance gain of the proposed algorithm in per-antenna rotation optimization, we consider two benchmark algorithms: 1) \textbf{two-layer RA with alternating rotation selection (two-layer-RA-ARS)}, where per-antenna rotation is optimized by a sequential linear search with the rotation angles of the other $N-1$ antennas being fixed; and 2) \textbf{two-layer RA with center-steering array rotation (two-layer-RA-CSAR)}, where boresights of all antennas are steered toward the center of the target region with per-antenna rotation being set as (14).

In Fig.~2, we compare the minimum beamforming gain of the proposed two-layer RA against three benchmarks across various target spatial regions. For a narrow region centered at broadside (Fig.~2(a)), the two-layer RA achieves approximately $97.8\%$ improvement over all benchmarks, which perform similarly. This is because array-only rotation misaligns antenna boresights with the target region, while the marginal variation in directional gain across the narrow region renders antenna-wise RA nearly equivalent to NRA. For a wide region centered at broadside (Fig.~2(b)), both antenna-wise and array-wise RAs yield only marginal gains over NRA, whereas the two-layer RA still achieves substantial improvement. For a narrow off-broadside region (Fig.~2(c)), antenna-wise rotation becomes the dominant factor, both antenna-wise RA and two-layer RA significantly outperform array-wise RA and NRA, improving the minimum beamforming gain by approximately $25.0\%$ and $48.7\%$, respectively. Finally, Fig.~2(d) shows that, given a general multi-region scenario, our two-layer RA can jointly adjust the rotation of the array and the antennas to form a beam pattern better adapted to the given region, achieving a higher improvement in minimum beamforming gain than other benchmarks.

\begin{figure}[!t]
 \centering
 \begin{subfigure}[t]{0.24\textwidth}
   \centering
   \includegraphics[width=1\textwidth]{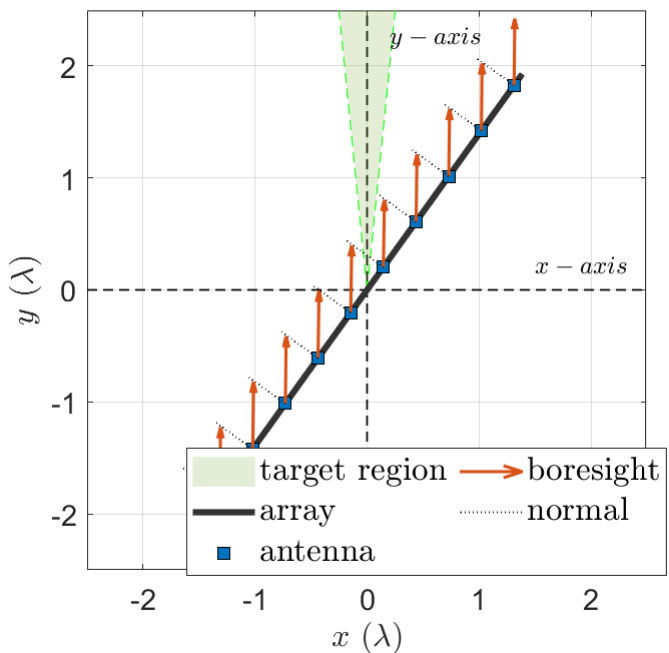}
   \caption{Target region $[-0.1,0.1]$}
   \label{fig:arc2a}
 \end{subfigure}%
 \hfill
 \begin{subfigure}[t]{0.24\textwidth}
   \centering
   \includegraphics[width=1\textwidth]{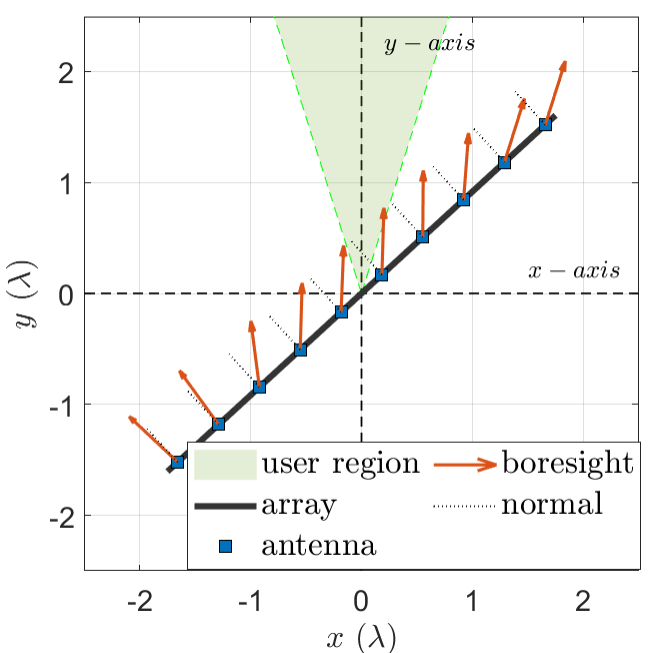}
   \caption{Target region $[-0.3,0.3]$}
   \label{fig:arc2b}
 \end{subfigure}%
 \hfill
 \caption{The optimized rotation of the proposed two-layer RA for different target spatial regions.}
 \label{fig:beam-patterns}
\end{figure}

Figs. 3(a) and 3(b) illustrate the optimized two-layer RA configurations for narrow and wide target regions, respectively. For the narrow region $[-0.1,0.1]$ (Fig.~3(a)), the array rotates off-broadside to widen the beam for better coverage. To compensate for scanning loss, the individual antenna boresights are steered toward the target center to retain peak gain. For the wide region $[-0.3,0.3]$ (Fig. 3(b)), a smaller array rotation mitigates gain drops at the boundaries, while individual antenna boresights disperse across the region to synthesize a flatter pattern and ensure uniform coverage.

\begin{figure}[!t]
 \centering
 \begin{minipage}[t]{0.24\textwidth}
  \centering
  \includegraphics[width=\textwidth]{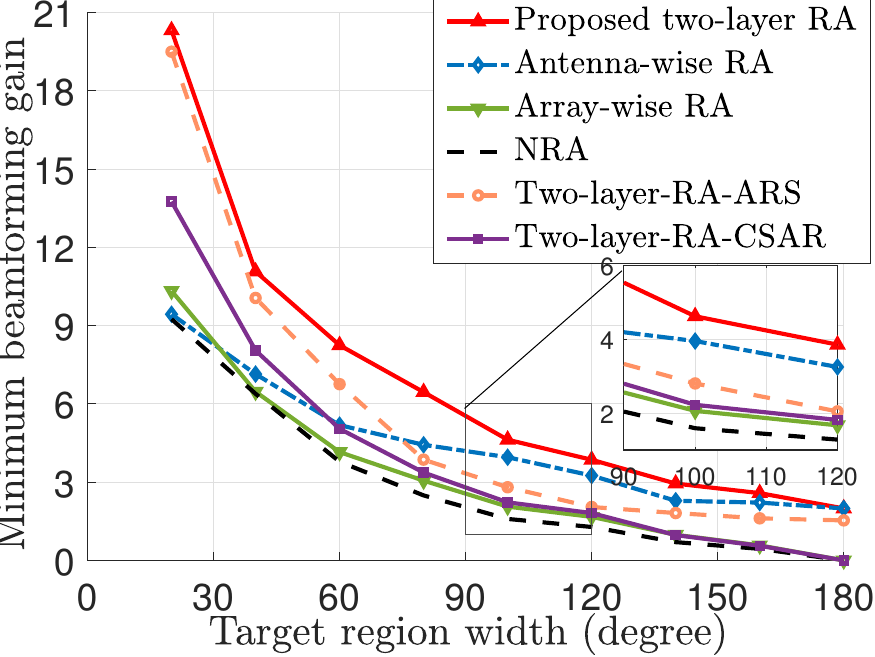}
  \caption{The minimum beamforming gain versus target angular region width $\theta_{\rm w}$.} 
  \label{fig:gain_vs_width}
 \end{minipage}
 \hfill
 \begin{minipage}[t]{0.24\textwidth}
  \centering
  \includegraphics[width=\textwidth]{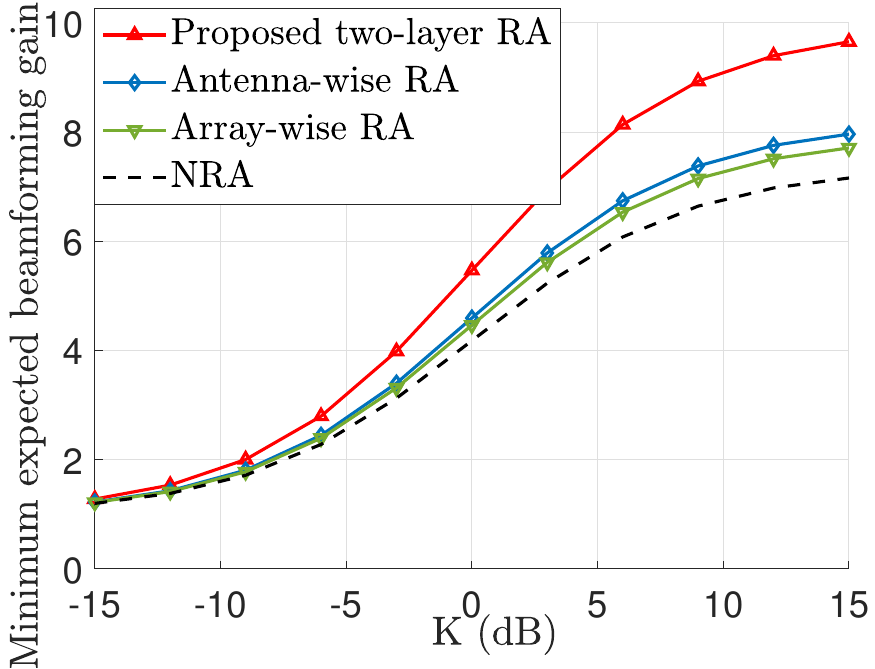}
  \caption{Minimum expected beamforming gain versus the Rician factor $K$.} 
  \label{fig:rician}
 \end{minipage}
 \vspace{-15pt}
\end{figure}

Fig. \ref{fig:gain_vs_width} compares the minimum beamforming gain versus the angular width $\theta_{\rm w}$, where the target region is $[-\theta_{\rm w}/2,\theta_{\rm w}/2]$. Two-layer RA consistently achieves the highest minimum beamforming gain, significantly outperforming two-layer-RA-ARS and two-layer-RA-CSAR, validating the proposed SCA-based optimization. Both antenna-wise RA and array-wise RA yield negligible improvement over NRA for narrow regions ($\theta_{\rm w} < 30^\circ$), whereas antenna-wise RA significantly enhances the minimum beamforming gain for wide regions ($\theta_{\rm w} > 80^\circ$). By synergistically exploiting both array-wise and antenna-wise DoFs, two-layer RA achieves robust performance across diverse coverage requirements.

Fig. \ref{fig:rician} shows the minimum expected beamforming gain versus the Rician factor $K$ for different benchmarks within an given angular region $[-0.3,0.3]$. It is observed that the two-layer RA yields significant gains in dominant LoS scenarios (e.g., when $K = 15$ dB), while the gain reduces when the Rician factor decreases.

\section{Conclusions}
In this letter, we have proposed a new two-layer RA architecture to enable flexible beamforming design for wireless communication systems. By jointly optimizing array and per-antenna rotation, the proposed two-layer RA exploits additional spatial DoFs to adaptively align both the antenna boresights and the array orientation toward the target regions. We formulated a max-min optimization problem to maximize the minimum beamforming gain within a target region and developed an efficient algorithm to solve it. Numerical results demonstrated that the proposed two-layer RA significantly outperforms conventional NRA and single-layer RA schemes, particularly in narrow-region and multi-region scenarios.

\begingroup
  \setlength{\itemsep}{-1pt}   
  \setlength{\parskip}{-3pt}   

\endgroup

%
\end{document}

%% file: header.tex
\newtheorem{remark}{\bf \emph{\underline{Remark}}}

\def\({\left(}
\def\){\right)}

\def\b0{{\mathbf{0}}}

